\documentclass[preprint,9.5pt]{elsarticle}

\usepackage{amssymb}

\journal{Mathematical Computation
}

\begin{document}

\begin{frontmatter}



\title{One generator $(1+u)$-quasi twisted codes over $F_2+uF_2$}


\author{Jian Gao$^1$, ~Qiong Kong $^2$}

\address{1.Chern Institute of Mathematics and LPMC, Nankai University, Tianjin~300071, China\\
         2. School of Science, Shandong University of Technology, Zibo, Shandong~255091, China }

\begin{abstract}
This paper gives the minimum generating sets of three types of one generator $(1+u)$-quasi twisted (QT) codes over $F_2+uF_2$, $u^2=0$. Moreover, it discusses the generating sets and the lower bounds on the minimum Lee distance of a special class of $A_2$ type one generator $(1+u)$-QT codes. Some good (optimal or suboptimal) linear codes over $F_2$ are obtained by these types of one generator $(1+u)$-QT codes.
\end{abstract}


\end{frontmatter}


\section{Introduction}
In recently, it has been shown that codes over finite rings are a very important class and many types of codes with good parameters could be constructed over rings [1,3,6,9]. Lately, there are some research on coding theory over finite chain ring $F_q+uF_q+\cdots +u^{s-1}F_q$, where $q$ is a positive power of some prime number $p$ and $s\geq 2$, the only finite chain ring with character $p$ [1-2,6-8]. In [8], Shi investigated the structural properties and the minimum generating sets of constacyclic codes over $F_q+uF_q+\cdots +u^{s-1}F_q$. Abualrub gave the classification of $(1+u)$-constacyclic codes with arbitrary length over $F_2+uF_2$ [2]. In [7], Kai studied the structural properties of $(1+\lambda u)$-constacyclic codes over $F_q+uF_q+\cdots +u^{s-1}F_q$.
\vskip 3mm Quasi-cyclic (QC) codes over commutative rings constitute a remarkable generalization of cyclic codes. More recently, they are produced many codes over finite fields which meet the best value of minimum distances of the same length and dimension [3,6,9]. Quasi-twisted (QT) codes as a generalization of QC codes, they also have some good algebraic structures and they are also produced many good codes over finite fields [4,5].
\vskip 3mm In this paper, we mainly research one generator $(1+u)$-QT codes over $F_2+uF_2$. The rest of the present paper organized as follows. In Sec.2, we survey some well known results related to our work. In Sec.3, we give the generating sets of the three types $(1+u)$-QT codes over $F_2+uF_2$. In Sec.4, we consider a special class of QT codes, which can lead to construct some good codes over $F_2$.

\section{Preliminaries}
\vskip 3mm Let $R=F_2+uF_2$, $u^2=0$, and $R^n=\{(c_0, c_1, \ldots, c_{n-1})|~c_i\in R, i=0,1,\ldots,n-1\}$. An nonempty set $C$ of $R^n$ is called a \emph{linear code} if and only if $C$ is an $R$-submodule of $R^n$. Defined a linear operator $T$ on $R^n$ such that $T(c)=((1+u)c_{n-1}, c_0, \ldots, c_{n-2})$ for each $c=(c_0, c_1, \ldots, c_{n-1})\in C$. A linear code $C$ is called an \emph{$(1+u)$-constacyclic code} of length $n$ over $R$ if and only if $T(C)=C$.
\vskip 3mm Let $S_n=R[x]/(x^n-(1+u))$. Define an $R$-module isomorphism as follows. $$\varphi:~R^n\rightarrow S_n$$ $$(c_0, c_1, \ldots, c_{n-1})\mapsto c_0+c_1x+\cdots +c_{n-1}x^{n-1}$$ One can verify that $C$ is an $(1+u)$-constacyclic code if and only if $\varphi (C)$ is an ideal of $S_n$. In this paper, we equal $(1+u)$-constacyclic code to the ideal of $S_n$.
\vskip 3mm The following two lemmas will be used in the discussing minimum generating sets of one generator $(1+u)$-QT codes.
\vskip 3mm \noindent {\bf Lemma 2.1 }(cf. [2] Lemma2)\emph{ Let $C$ be an $(1+u)$-constacyclic code of length $n$ over $R$. Then $C=( g(x)+up(x), ua(x))$, where $a(x)|g(x)|(x^n-1)$ over $F_2$ and ${\rm deg}a(x)> {\rm deg}p(x)$.}

\vskip 3mm \noindent {\bf Lemma 2.2 }(cf. [2] Corollary 1) \emph{Let $C=( g(x)+up(x), ua(x))$ be an $(1+u)$-constacyclic code of length $n$ over $R$, where $n=2^em$, ${\rm gcd}(2,m)=1$. Then $C$ has the following three types:}
\vskip 1mm \emph{(1)~$C=(g(x))$, where $g(x)|(x^n-1)$ over $F_2$;}
\vskip 1mm \emph{(2)~$C=(ug(x))$, where $g(x)|(x^n-1)$ over $F_2$;}
\vskip 1mm (\emph{3)~$C=(f_1^{i_1}\cdots f_r^{i_r})$, where $f_i|(x^n-1)$ over $F_2$ and there exists an $i_j$ such that $2^e\leq i_j\leq 2^{e+1}$.}

\vskip 3mm We call the above three types $(1+u)$-constacyclic codes $A_1$ type, $A_2$ type and $B$ type, respectively.

\vskip 3mm Define the Lee weight $W_L$ of the elements $0, 1, u, 1+u$ as $0, 1, 2, 1$, respectively. Moreover, the Lee weight of an $n$-tuple in $R^n$ is the sum of the Lee weights of its components. The Gray map $\phi$ sends the elements $0,1,u,1+u$ of $R$ to $(0,0), (0,1), (1,1), (1,0)$ over $F_2$, respectively. It is easy to verify that $\phi$ is a linear isometry form $R^n$ (Lee distance) to $F_2^{2n}$ (Hamming distance).
\section{One generator $(1+u)$-QT codes}
 A linear code $C$ is called \emph{quasi-twisted} (QT) code if it is invariant under $T^l$ for some positive integer $l$. The smallest $l$ such that $T^l(C)=C$ is called the index of $C$. Clearly, $l$ is a divisor of $N$. Let $N=nl$. Define a one-to-one correspondence $$\rho: R^N\rightarrow S_n^l$$ $$(a_{00}, a_{01}, \ldots, a_{0,l-1};a_{10}, a_{11}, \ldots, a_{1,l-1}; \ldots; a_{n-1,0}, a_{n-1,1}, \ldots, a_{n-1,l-1})$$$$\mapsto  \underline{f}(x)=(f_0(x), f_1(x), \ldots,  f_{l-1}(x))$$ where $f_j(x)=\sum_{i=0}^{n-1}a_{ij}x^i$ for $j=0,1,\ldots,l-1$. Then $C$ is equivalent to for any $\underline{f}(x)=(f_0(x), f_1(x), \ldots, f_{l-1})\in \rho(C)$, $x\underline{f}(x)\in \rho(C)$. Therefore, $C$ is a QT code if and only if $\rho(C)$ is an $R[x]$-submodule of $S_n^l$. Let $C=S_n\underline{f}(x)$, where $\underline{f}(x)$ is defined as above. Then $C$ is called a\emph{ one generator $(1+u)$-QT code}. For simplicity, we denote $C=(f_0(x), f_1(x), \ldots, f_{l-1}(x))$.

\vskip 3mm Let $C$ be a one generator $(1+u)$-QT code of length $nl$ with index $l$ over $R$, where $n=2^em$, ${\rm gcd}(2,m)=1$. Assume that $C=(f_0(x), f_1(x), \ldots,f_{l-1}(x))$, where $f_j(x)\in S_n$, $j=0,1,\ldots,l-1$. For each $i=0,1,\ldots,l-1$, define an $R[x]$-module homomorphism as follows $$\psi_i:~S_n^l\rightarrow S_n$$ $$(f_0(x), f_1(x), \ldots, f_{l-1}(x))\mapsto f_i(x)$$ Then $\psi_i(C)$ is an ideal in $S_n$, i.e., $\psi_i(C)$ is an $(1+u)$-constacyclic code of length $n$ over $R$. If $\psi_i(C)$ is the type of $A_1$, $A_2$ or $B$, then we call $C$ is the $A_1$, $A_2$ or $B$ type one generator $(1+u)$-QT code, respectively.

\vskip 3mm \noindent {\bf Theorem 3.1} \emph{Let $C$ be an $A_1$ type one generator $(1+u)$-QT code of length $nl$ with index $l$ generated by $G=(g_0, g_1, \ldots, g_{l-1})$ over $R$, where $n=2^em$, ${\rm gcd}(2,m)=1$ and $g_i\in F_2[x]$, $i=0,1,\ldots,l-1$. Let $g={\rm gcd}(g_0, g_1, \ldots, g_{l-1}, x^n-1)$, $h=(x^n-1)/g$, ${\rm deg}h=r$, $f_i=g_i/g$, $F=\{uf_0, uf_1, \ldots, uf_{l-1}\}$. Then the minimum generating set of $C$ is $S_1\cup S_2$, where $S_1=\{G, xG, \ldots, x^{r-1}G\}$ and $S_2=\{F, xF, \ldots, x^{n-r-1}F\}$.}
\vskip 1mm \noindent\emph{ Proof} Let $c(x)$ be a codeword of $C$. Then there exists $f(x)\in S_n$ such that
\begin{equation}
c(x)=f(x)G=f(x)(g_0, g_1, \ldots, g_{l-1})
\end{equation}
Using Euclidean division, there are polynomials $Q_1, R_1\in R[x]$ such that
\begin{equation}
f(x)=Q_1h+R_1
\end{equation}
where ${\rm deg}Q_1\leq n-k-1$, $R_1=0$ or ${\rm deg}R_1\leq r-1$. Therefore from $(1)$ and $(2)$, we have $$c(x)=Q_1h(g_0, g_1, \ldots, g_{l-1})+R_1(g_0, g_1, \ldots, g_{l-1})$$ Clearly, $R_1(g_0, g_1, \ldots, g_{l-1})\in {\rm Span}(S_1)$. Since $(g_0, g_1, \ldots, g_{l-1})=g(f_0, f_1, \ldots, f_{l-1})$, we have
\begin{equation}
Q_1h(g_0, g_1, \ldots, g_{l-1})=Q_1hg(f_0, f_1, \ldots, f_{l-1})
\end{equation}
Note that $u=x^n-1=gh$ in $S_n$. Therefore $(3)=Q_1(uf_0, uf_1, \ldots, uf_{l-1})$, i.e., $$Q_1h(g_0, g_1, \ldots, g_{l-1})\in {\rm Span}(S_2)$$ which implies that $S_1\cup S_2$ generates $C$. Next, we will prove ${\rm Span}(S_1)\cap {\rm Span}(S_2)=\{0\}$.
\vskip 1mm Let $e(x)=(e_0(x), e_1(x), \ldots, e_{l-1}(x))\in {\rm Span}(S_1)\cap {\rm Span}(S_2)$. Since $e(x)\in {\rm Span}(S_1)$, it follows that
\begin{equation}
e_j(x)=g_j(\alpha_0+\alpha_1x+\cdots +\alpha_{r-1}x^{r-1})~\forall j=0,1,\ldots,l-1
\end{equation}
On the other hand, since $e(x) \in {\rm Span}(S_2)$,
\begin{equation}
e_j(x)=uf_j(\beta_0+\beta_1x+\cdots +\beta_{n-r-1}x^{n-r-1})~\forall j=0,1,\ldots,l-1
\end{equation}
From $(5)$, for each $j=0,1,\ldots,l-1$, $ue_j=0$, which implies that $\alpha_i=0$ or $\alpha_i=u$, $i=0,1,\ldots,r-1$. Assume that $$M_1(x)=\alpha_0+\alpha_1x+\cdots+\alpha_(r-1)x^{r-1}$$ $$M_2(x)=\beta_0+\beta_1x+\cdots+\beta_{n-r-1}x^{n-r-1}$$ Then $g_jM_1(x)=uf_jM_2(x)$. Thus $ug_j(M_1(x)+hM_2(x))=2u^2f_jM_2(x)=0$. Using the facts $e_j(x)\in {\rm Span}(S_1)$ and $M_2(x)\in F_2[x]$, we have that $M_1(x)+hM_2(x)=0$. SO for each $i=0,1,\ldots, r-1$ and $j=0,1,\ldots,n-r-1$, $\alpha_i=\beta_j=0$. It means that ${\rm Span}(S_1)\cap {\rm Span}(S_2)=\{0\}$.                 \hfill $\Box$

\vskip 3mm \noindent {\bf Theorem 3.2} \emph{Let $C$ be an $A_2$ type one generator $(1+u)$-QT code of length $nl$ with index $l$ generated by $G=(ug_0, ug_1, \ldots, ug_{l-1})$ over $R$, where $n=2^em$, ${\rm gcd}(2,m)=1$, $g_i(x)\in F_2[x]$, $i=0,1,\ldots,l-1$. Let $g={\rm gcd}(g_0, g_1, \ldots, g_{l-1}, x^n-1)$, $h=(x^n-1)/g$, ${\rm deg}h=r$. Then the minimum generating set of $C$ is $S=\{G, xG, \ldots, x^{r-1}G\}$. }
\vskip 1mm \noindent\emph{ Proof} Let $c(x)$ be a codeword of $C$. Then there exists $f(x)\in S_n$ such that
\begin{equation}
c(x)=f(x)G=f(x)(ug_0, ug_1, \ldots, ug_{l-1})
\end{equation}
Using Euclidean division, there are polynomials $Q_1, R_1\in R[x]$ such that
\begin{equation}
f(x)=Q_1h+R_1
\end{equation}
where $R_1=0$ or ${\rm deg}R_1\leq r-1$. Therefore from $(6)$ and $(7)$, we have $$c(x)=uQ_1h(g_0, g_1, \ldots, g_{l-1})+R_1(ug_0, ug_1, \ldots, ug_{l-1})$$ Note that $u=x^n-1=gh$ and $g|g_i$ in $S_n$, $i=0,1,\ldots,l-1$. Therefore $$c(x)=R_1(ug_0, ug_1, \ldots, ug_{l-1}),$$ which implies that $S$ generates $C$. By the construction of $S$, $S$ is minimum. Thus $S$ is the minimum generating set of $C$.               \hfill $\Box$

\vskip 3mm \noindent {\bf Theorem 3.3} \emph{Let $C$ be an $B$ type one generator $(1+u)$-QT code of length $nl$ with index $l$ generated by $G=(q_0fg, q_1fg, \ldots, q_{l-1}fg)$ over $R$, where $n=2^em$, ${\rm gcd}(2,m)=1$, $q_i\in F_2[x]$, $i=0,1,\ldots,l-1$, and $g\in F_2[x]$ with the maximal degree satisfying $f|g|(x^n-1)$. Let $h=(x^n-1)/g$, ${\rm deg }g=r$, ${\rm deg} f=t$, ${\rm gcd}(q_i, h)=1$ and $F=(uq_0f, uq_1f, \ldots, uq_{l-1}f)$. Then the minimum generating set of $C$ is $S_1\cup S_2$, where $S_1=\{G, xG, \ldots, x^{n-r-1}G\}$ and $S_2=\{F, xF, \ldots, x^{r-t-1}F\}$.}
\vskip 1mm \noindent\emph{ Proof} We just prove ${\rm Span}(S_1)\cap {\rm Span}(S_2)=\{0\}$.
\vskip 1mm Let $e(x)=(e_0(x), e_1(x), \ldots, e_{l-1}(x))\in {\rm Span}(S_1)\cap {\rm Span}(S_2)$. Since $e(x)\in {\rm Span}(S_1)$, it follows that
\begin{equation}
e_j(x)=q_jfg(\alpha_0+\alpha_1x+\cdots +\alpha_{n-r-1}x^{n-r-1})~\forall j=0,1,\ldots,l-1
\end{equation}
On the other hand, since $e(x) \in {\rm Span}(S_2)$,
\begin{equation}
e_j(x)=uq_jf(\beta_0+\beta_1x+\cdots +\beta_{r-t-1}x^{r-t-1})~\forall j=0,1,\ldots,l-1
\end{equation}
From $(9)$, for each $j=0,1,\ldots,l-1$, $ue_j=0$, which implies that $\alpha_i=0$ or $\alpha_i=u$, $i=0,1,\ldots,r-1$. Assume that $$M_1(x)=\alpha_0+\alpha_1x+\cdots+\alpha_{n-r-1}x^{n-r-1}$$ $$M_2(x)=\beta_0+\beta_1x+\cdots+\beta_{r-t-1}x^{r-t-1}$$ Then $q_jfgM_1(x)=uq_jfM_2(x)$. Thus $uq_jfg(M_1(x)+hM_2(x))=2u^2f_jM_2(x)=0$, which implies that $q_jfg|(x^n-1)$ or $M_1(x)+hM_2(x)=0$. Since ${\rm gcd}(q_i, h)=1$, it follows that $q_jfg$ does not divide $x^n-1$. Therefore $M_1(x)+hM_2(x)=0$. So for each $i=0,1,\ldots,n-r-1$ and $j=0,1,\ldots,r-t-1$, we have $\alpha_i=\beta_j=0$, i.e., ${\rm Span}(S_1)\cap {\rm Span}(S_2)=\{0\}$.         \hfill $\Box$

\vskip 3mm  In the rest of this section, we present some examples to illustrate the applications to these theorems.

\vskip 3mm \noindent {\bf Example 3.4}
\vskip 1mm $\bullet$~Taking $n=3$, $l=2$ with $g_0=x+1$ and $g_1=x^2+1$, we get an $A_1$ type one generator $(1+u)$-QT code $C$ of length $3\times2=6$ with index $2$ over $R$. Since $g=x+1$, $f_0=1$ and $f_1=x+1$, from Theorem 3.1 the generating set of $C$ is $\{(g_0, g_1), x(g_0,g_1)\}\cup \{(uf_0, uf_1)\}$, which imlies that $|C|=4^22$. By the Gray map, we get $\phi(C)$ is an optimal $[12, 5 ,4]$ linear code over $F_2$.
\vskip 1mm $\bullet$~Taking $n=3$, $l=3$ with $g_0=x^4+x^3+x^2+1$ and $g_1=x^4+x^3+x+1$, $g=x+1$, $f_0=x^3+x+1$, $f_1=x^3+x^2+1$ and $f_2=1$, we can get an $A_2$ type one generator $(1+u)$-QT code $C$ of length $9$ with index $3$ over $R$. Since $g=x+1$, from Theorem 3.2 the generating set of is $\{(ug_0, ug_1), x(ug_0, ug_1)\}$, which implies that $|C|=2^2$ over $R$. By the Gray map, $\phi(C)$ is an optimal $[18,2,12]$ linear code over $F_2$.
\vskip 1mm $\bullet$~Taking $n=9$, $l=2$ with $f=x+1$, $g=(x+1)(x^6+x^3+1)$, $q_0=x$ and $q_1=x+x^2$, we get an $B$ type one generator $(1+u)$-QT code $C$ of length $9\times2=18$ with index $2$ over $R$. From Theorem 3.3, we have the generating set of $C$ is $\{(q_0fg, q_1fg), x(q_0fg, q_1fg)\}\cup \{(uq_0fg, uq_1fg), \ldots, x^5(uq_0fg, uq_1fg)\}$, which implies that $|C|=4^22^6$. By the Gray map, we get $\phi(C)$ is an $[36,10,8]$ linear code over $F_2$.
\section{A special $A_2$ type}
 In this section, we discuss a special class of $A_2$ type one generator $(1+u)$-QT codes. We determine the minimum generating sets of them and introduce a lower bound for the minimum Lee distance.

\vskip 3mm \noindent {\bf Theorem 4.1}\emph{ Let $C$ be an $A_2$ type one generator $(1+u)$-QT code of length $nl$ with index $l$ generated by $G=(ugf_0, ugf_1, \ldots, ugf_{l-1})$ over $R$, where $n=2^em$, ${\rm gcd}(2,m)=1$, $g,f_i\in F_2[x]$, $g|(x^n-1)$ and ${\rm gcd}(f_i, (x^n-1)/g)=1$, $i=0,1,\ldots,l-1$. Let ${\rm deg}g=n-r$. Then the minimum generating set of $C$ is $G=\{G, xG, \ldots, x^{r-1}G\}$. Moreover, the minimum Lee distance of $C$ is $d_L(C)\geq ld_L(\widetilde{C})$, where $\widetilde{C}=(ug)$.}
\vskip 1mm \noindent\emph{ Proof} From Theorem 3.2, one can verify that the minimum generating set of $C$ is $G=\{G, xG, \ldots, x^{r-1}G\}$. On the other hand, if for each $i=0,1,\ldots,l-1$ ${\rm gcd}(f_i, (x^n-1)/g)=1$, then $(ug)=(ugf_i)$. It follows that $\psi_i(C)=\widetilde{C}$ for each $i=0,1,\ldots,l-1$. Let $c(x)=(ugf_0a(x), ugf_1a(x), \ldots, ugf_{l-1}a(x))$ be a codeword of $C$. Then $ugf_ia(x)=0$ if and only if $(x^n-1)|gf_ia(x)$ if and only if $gh|gf_ia(x)$ if and only if $h|f_ia(x)$ if and only if $h|a(x)$. It means that if $ugf_ia(x)=0$, then $c(x)=0$, $i=0,1,\ldots,l-1$. Therefore for any nonzero codeword $c(x)$, $\psi_i(c(x))\neq 0$, which implies that $d_L(C)\geq ld_L(\widetilde{C})$.                 \hfill $\Box$

\vskip 3mm Using Theorem 4.1, some examples of this family that yield good (optimal or suboptimal) linear codes are given.

\vskip 3mm \noindent {\bf Example 4.2}
\vskip 1mm $\bullet$~Taking $n=3$, $l=2$ with $g=x+1$, $f_0=x^3+x+1$ and $f_1=x^3+x^2+1$, we get an $A_2$ type one generator $(1+u)$-QT code $C$ of length $3\times2=6$ with index $2$ over $R$. From Theorem 4.1, $|C|=2^{3-1}=4$ and $d_L(C)\geq 2\times 4=8$. Since there are some codewords with Lee weight $8$, the minimum Lee weight of $C$ is $8$ actually, i.e., $C$ is an $(6, 4, 8)_L$ linear code over $R$. By the Gray map, we get $\phi(C)$ is an $[12, 2 ,8]$ linear code, which is optimal over $F_2$.
\vskip 1mm $\bullet$~Taking $n=4$, $l=2$ with $g=x+1$, $f_0=x^4+x+1$ and $f_1=x^2+x+1$, by the help of Theorem 4.1 and the Gray map, we have $\phi(C)$ is an suboptimal $[16,3,8]$ linear code over $F_2$.
\vskip 1mm $\bullet$~Taking $n=4$, $l=3$ with $g=x+1$, $f_0=x^4+x+1$, $f_1=x^2+x+1$ and $f_2=x^2$, by the help of Theorem 4.1 and the Gray map, we have $\phi(C)$ is an optimal $[24,3,12]$ linear code over $F_2$.

\vskip 3mm \noindent {\bf Acknowledgments} This research is supported in part by the Natural Science Foundation of Shandong provence (Grant No. ZR2011AQ004).


\end{document}